# Design of Advanced Readout and System-on-Chip Analog Circuits for Quantum Chip

Ahmad Salmanogli[1,3], Hesam Zandi[2,3,6], Mahdi Esmaeili Turan Poshti[3], M. Hossein Eskandari[3], Ertan Zencir[4], Mohsen Akbari[5]

[1]Department of Electrical and Electronics, Engineering Faculty, Ankara Yildirim Beyazit University, Turkey
[2]Faculty of Electrical Engineering, K. N. Toosi University of Technology, Tehran, Iran
[3]Iranian Quantum Technologies Research Center (IQTEC), Tehran, Iran
[4]Department of Electrical and Electronics Engineering, University of Turkish Aeronautical, Ankara, Turkey
[5]Quantum optics lab, Department of Physics, Kharazmi University, Tehran, Iran
[6]Electronic Materials Laboratory, K. N. Toosi University of Technology, Tehran, Iran

**Abstarct:** In this work, we design an advanced quantum readout architecture that integrates a four-qubit superconducting chip with a novel parametric amplifier ended with analog front-end circuit. Unlike conventional approaches, this design eliminates the need for components such as Purcell filters. Instead, a Josephson Parametric Amplifier is engineered to simultaneously perform quantum-limited signal amplification and suppress qubit energy leakage. The design features a tailored gain profile across C-band, with sharp peaks (~24 dB) and troughs (~0 dB), enabling qubit frequencies to align with gain minima and resonator frequencies with gain maxima. By leveraging strategic frequency placement, a simplified and efficient quantum chip architecture is proposed and designed. The designed quantum chip is modeled using Qiskit Metal and further analyzed through a quantum theoretical framework, which reveals the emergence of entanglement in the current probabilities of coupled qubits. Complementing the quantum hardware, a 45 nm CMOS System-on-Chip incorporating essential modules is developed. Post-layout simulations demonstrate wideband operation (4.1–8.2 GHz), high gain (~72 dB), low noise (<1 dB), compact footprint (0.6×0.4 mm²), and low power (~70 mW). The design traditionally can support up to 160 qubits, with a per-qubit power budget of approximately 0.44 mW and offers a scalable quantum systems.

**Key words:** Quantum chip, Entanglement, JPA, SoC, 45 nm CMOS

## Introduction

Designing high-performance quantum chips is a multifaceted task involving qubit coherence, interconnectivity, error correction, and scalability. Superconducting qubits, particularly transmons, are preferred due to their fabrication compatibility and extended coherence times. However, performance can be degraded by crosstalk, decoherence, and fabrication flaws [1–7]. Recent developments include automated coupler design methodologies to optimize qubit-coupler-qubit setups, boosting scalability and performance [2]. Additionally, the layout and interconnectivity of qubits significantly affect system behavior. A comprehensive study found that increased connectivity generally improves performance, regardless of specific topology [3]. Therefore, enhancing quantum chip efficiency requires a holistic approach: refining qubit design and layout, incorporating robust error correction, and carefully managing interconnectivity. Key performance metrics include coherence times, frequency tunability, readout fidelity, and suppression of thermal and magnetic noise. Electromagnetic design is vital for simulating coupling strengths, frequencies, and parasitic modes, while materials engineering such as substrate selection and surface treatments helps minimize two-level system noise [8,9]. To ensure high-fidelity readout without compromising the critical factors mentioned, quantum-limited amplifiers such as Josephson Parametric

Amplifiers (JPAs) [15–22] and Purcell filters [10–14] should be carefully designed and integrated into quantum systems. Qubits are usually linked to readout resonators that connect to a measurement line, which may inadvertently allow energy leakage—known as Purcell decay. To counter this, Purcell filters are inserted between the resonator and the readout chain to block qubit-frequency signals while passing the readout signals, enhancing measurement fidelity [10–14]. On the amplification side, JPAs use the nonlinear properties of Josephson junctions (JJ) to amplify weak qubit signals with minimal added noise. Unlike conventional amplifiers, JPAs approach the quantum noise limit and are typically used as the first amplification stage, followed by HEMT or CMOS-based amplifiers [23–29]. In addition to improving signal detection, JPAs enable advanced experiments such as quantum squeezing and entanglement measurement, underscoring their essential role in quantum technologies [15–22]. Nonetheless, in this study, an advanced design is introduced based on the Blochnium Josephson Parametric Amplifier (BJPA) [22], to exhibit a unique gain profile suited for quantum applications. This tailored gain spectrum matches the quantum chip's frequency layout, eliminating the need for Purcell filters. The introduced JPA not only amplifies signals but also mitigates qubit energy leakage and enhance readout efficiency due to reducing in quantum circuit complexity.

Post-JPA amplification, ultra-weak microwave signals generated by quantum chip are further processed through an analog System-on-Chip (SoC). The SoC includes critical components such as a Low Noise Amplifier (LNA), Voltage-Controlled Oscillator (VCO), mixer, RF/IF amplifiers, and bandpass filters. The LNA is especially crucial, as it determines the overall noise figure of the receiver and must maintain ultra-low noise levels to preserve the quantum signal's integrity [23–29, 30]. High RF gain in the SoC ensures signals surpass the noise floor, reducing the need for multiple discrete amplification stages and improving stability [31]. The SoC compactness and low power are other crucial factors for integration in quantum chips. Along with SoC smaller die size reduces parasitic effects and supports better matching and integration density [32], while low power operation minimizes thermal load—vital for maintaining qubit coherence [9]. Thus, the analog SoC serves as a compact, low-noise bridge between quantum hardware and classical electronics, essential for scalable quantum computing.

In this work, we propose a novel readout architecture and an analog front-end circuit tailored for a four-qubit quantum chip illustrated in scheme 1. The signals generated by the quantum chip, around –135 dBm, are amplified by a JPA, Purcell filter involved, to reach detectable levels (~–110- –105 dBm). The amplified signal is then processed at room temperature by a SoC, which includes RF and IF components altogether to extract the zero-IF quantum information efficiently. By engineering the JPA with a novel gain profile, the need for conventional Purcell filters is eliminated. This configuration significantly suppresses qubit relaxation while simultaneously maximizing readout signal amplification. As another main and complementary task, the quantum chip is also modeled using Qiskit Metal and analyzed within a rigorous quantum theoretical framework by which the entanglement properties generated is investigated. Complementing this, a 45 nm CMOS-based SoC analog receiver is designed, featuring a planar architecture that delivers 72 dB RF gain, a noise figure below 1 dB, and a low power footprint of approximately 70 mW. The SoC also supports a wide operational bandwidth ranging from 4.1 to 8.2 GHz and occupies a compact silicon footprint of just 0.6×0.4 mm², making it highly suitable for scalable quantum systems. In the following sections, we begin with a theoretical analysis of the individual subsystems, shown in scheme 1, which is a schematic overview of the quantum chip, a multifunctional JPA, followed by a versatile custom-designed SoC module. That is followed by a detailed step-by-step design and examination of each component, culminating in a presentation of the results.

## Theoretical background

The quantum chip design is fundamentally guided by the unique gain profile of the JPA. To rigorously examine the new JPA, the configuration—illustrated in Fig. 1—is subjected to a quantum mechanical analysis. This begins with the formal derivation of the system's Lagrangian and its corresponding total Hamiltonian. Subsequently, the dynamical behavior is explored using the quantum Langevin formalism, providing insight into signal amplification and energy leakage mechanisms. The total Lagrangian [18, 22] of the quantum circuit proposed just for i$^{th}$ cell, containing two successive Quartons coupled through a typical resonator, indicated in the figure is expressed as:

$$L_t = \frac{C_g}{2}\sum_{\substack{k=0\\k\neq 4}}^{8}\dot{\phi}_k^2 + \frac{C_J}{2}\sum_{k=0}^{7}\left(\dot{\phi}_k - \dot{\phi}_{k+1}\right)^2 + \frac{C_m}{2}\left\{\left(\dot{\phi}_0 - \dot{\phi}_4\right)^2 + \left(\dot{\phi}_4 - \dot{\phi}_8\right)^2\right\} + \left(\frac{C_1+C_2+C_s}{2}\right)\dot{\phi}_4^2 - \frac{1}{2L_s}\phi_4^2$$
$$-\frac{1}{2L_{Js}}\sum_{k=0}^{7}\left(\phi_k - \phi_{k+1}\right)^2 - \frac{1}{2L_{Jm}}\left\{\left(\phi_0 - \phi_4\right)^2 + \left(\phi_4 - \phi_8\right)^2\right\} + 8N*E_{Js}\cos\left(\frac{\varphi}{8N}\right) + N*E_{Jm}\cos\left(\frac{\varphi}{N}\right)$$

(1)

where $C_g$, ($C_{Js}$, $L_{Js}$, $E_{Js}$), ($C_{Jm}$, $L_{Jm}$, $E_{Jm}$), N, and $\phi$ are the parasitic capacitors grounded the SQUIDs, the secondary JJ capacitance inductance, and Josephson energy, the primary JJ capacitance, inductance, and Josephson energy, the number of primary SQUIDs in each cell, and flux as a quantum coordinate operator, respectively. To simplify the algebra, it is convenient to represent Eq. 1 through coupled equations. In this simplification approach, the last two terms in the Lagrangian are initially put aside, however they will be added while calculating the total Hamiltonian. Henceforth, one can express Eq. 1 as a compact form $L_t = \frac{1}{2}\dot{\phi}^T\hat{C}\dot{\phi} - \frac{1}{2}\phi^T\hat{L}^{-1}\phi$, [18, 19, 21-22] where $\hat{C}$ and $\hat{L}^{-1}$ are matrixes as:

$$\hat{C} = \begin{bmatrix} C_g+C_j+C_M & -C_j & 0 & 0 & -C_m & 0 & 0 & 0 & 0 \\ -C_j & C_g+2C_j & -C_j & 0 & 0 & 0 & 0 & 0 & 0 \\ 0 & -C_j & C_g+2C_j & -C_j & 0 & 0 & 0 & 0 & 0 \\ 0 & 0 & -C_j & C_g+2C_j & -C_j & 0 & 0 & 0 & 0 \\ -C_m & 0 & 0 & -C_j & C_X & -C_j & 0 & 0 & -C_m \\ 0 & 0 & 0 & 0 & -C_j & C_g+2C_j & -C_j & 0 & 0 \\ 0 & 0 & 0 & 0 & 0 & -C_j & C_g+2C_j & -C_j & 0 \\ 0 & 0 & 0 & 0 & 0 & 0 & -C_j & C_g+2C_j & -C_j \\ 0 & 0 & 0 & 0 & -C_m & 0 & 0 & -C_j & C_g+C_j+C_M \end{bmatrix}$$

$$\hat{L}^{-1} = L^{-1}\begin{bmatrix} 2 & -1 & 0 & 0 & -1 & 0 & 0 & 0 & 0 \\ -1 & 2 & -1 & 0 & 0 & 0 & 0 & 0 & 0 \\ 0 & -1 & 2 & -1 & 0 & 0 & 0 & 0 & 0 \\ 0 & 0 & -1 & 2 & -1 & 0 & 0 & 0 & 0 \\ -1 & 0 & 0 & -1 & L_X & -1 & 0 & 0 & -1 \\ 0 & 0 & 0 & 0 & -1 & 2 & -1 & 0 & 0 \\ 0 & 0 & 0 & 0 & 0 & -1 & 2 & -1 & 0 \\ 0 & 0 & 0 & 0 & 0 & 0 & -1 & 2 & -1 \\ 0 & 0 & 0 & 0 & -1 & 0 & 0 & -1 & 2 \end{bmatrix}$$

(2)

where $C_X = C_1 + C_2 + C_s + 2C_m + 2C_j$ and $L_X = 4 + L/L_s$. From Eq. 2, it is evident that the structure of the C and L matrices differs from that of the BJPA [22], and also $C_x$ and $L_x$ play central roles. This means that the dynamics applied by this new structure operating as a parametric amplifier should be differed than the BJPA [22]. Additionally, the new design introduces several extra degrees of freedom compared to the

BJPA, enabling the development of a more versatile and tunable JPA. These key advancements will be discussed in detail in the following sections. Using the matrices for capacitors and inductors presented in Eq. 2, we can define the matrix $\Omega^2 = \hat{C}^{-1}\hat{L}^{-1}$ [18, 19, 21] to calculate the eigenvalues (dominant frequencies) and eigenvectors ($\Psi_i$ is the wave profile of each mode) of the quantum system under discussion. Using these equations, one can calculate the effective capacitance and inductance related to the structure. The effective capacitance and inductance of the circuits can be determined using the eigenvalues and eigenvectors for each mode as follows: $C_{eff} = \vec{\Psi}_l^T \hat{C} \vec{\Psi}_l$ and $L_{eff}^{-1} = \vec{\Psi}_l^T \hat{L}^{-1} \vec{\Psi}_l$. This allows us to map a λ/4 resonator to an equivalent LC circuit [18, 22]. Consequently, the Blochnium structure designed in this work is modeled as a series combination of an effective inductors and capacitors, along with nonlinear elements due to the terms separated in Eq. 1. Analyzing this simplified circuit model enables to determine much more easily and effectively the quantum circuit impedance $Z_{eff} = \sqrt{L_{eff}/C_{eff}}$ and subsequently obtain the coupling rate of the quantum system to the environment $\kappa_{eff} = \omega_{eff}/Q_{eff}$, where $Q_{eff}$ is the circuit quality factor arisen due to the mismatching. In line with the approach taken in BJPA [22], the total Hamiltonian of the effective circuit [18, 22] and its associated dynamical behavior can be explored using quantum formalisms such as the quantum Langevin equation. This framework enables the calculation of both signal and idler gain for the proposed quantum circuit. Due to the extensive analytical derivations involved, readers are encouraged to consult reference [22] for a detailed treatment. It will be demonstrated that the gain spectrum significantly differs from that of the BJPA. This deviation likely arises from the distinct capacitance and inductance matrices (C and L) associated with the new architecture—specifically, the coupling of two Quartons via a resonator—which gives rise to fundamentally different system dynamics. However, like BJPA [22], the incorporation of both primary and secondary SQUIDs provides enhanced control over the circuit's nonlinearity and tunability, allowing for greater selectivity and amplified response. Following the design of a tailored JPA with a distinctive gain spectrum, and using its unique gain spectrum a quantum chip comprising four coupled qubits is developed. The specifics of the quantum system under investigation will be elaborated in the subsequent section. In this context, we conduct a detailed quantum theoretical analysis of the chip. The total Hamiltonian corresponding to the quantum architecture depicted in Fig. 3, for just coupling between qubit 2 ($Q_2$) and qubit 4 ($Q_4$) (it can be applied in the same way for $Q_1$ and $Q_3$) through the contributed bus resonator, is given by:

$$H_{Q2-Q4} = \hbar\omega_1 a_1^+ a_1 + \hbar\omega_2 a_2^+ a_2 + \hbar\omega_c b^+ b + \hbar g_1 (a_1^+ + a_1)(b^+ + b) + \hbar g_2 (a_2^+ + a_2)(b^+ + b) + \hbar g_3 (a_1^+ + a_1)(a_2^+ + a_2)$$
$$+ \hbar g (a_1^+ + a_1)(a_2^+ + a_2)(b^+ + b) \qquad (3)$$

where ($a_1^+$, $a_1$), ($a_2^+$, $a_2$), and ($b^+$, b) denote the creation and annihilation operators for qubit 2, qubit 4, and the central bus resonator, respectively. Also, $g_1$, $g_2$, $g_3$, and g are the coupling between $Q_2$ and the bus, the coupling between $Q_4$ and the bus, the coupling between $Q_2$ and the $Q_4$, and the coupling between $Q_2$, the bus, and $Q_4$, respectively. The first three terms of the Hamiltonian represent the energy contributions of $Q_2$, $Q_4$, and the bus resonator. The remaining terms describe the interaction and coupling dynamics between the qubits and the bus resonator. It is well understood that the JPA is responsible for sensing the output signals from the quantum chip, which typically arise as current fluctuations generated by the quantum dynamics of the system. Focusing specifically on Eq. 3, which describes the interaction between two qubits mediated by the bus resonator, the readout ultimately captures current signals induced by this qubit-bus-qubit coupling. These currents are directly proportional to the probability amplitudes of each qubit occupying its excited state, independent of the other's state [33, 34]. By analyzing the time evolution of these excitation probabilities—obtained by solving the system's quantum master equation [35-37]—it becomes possible to derive the time-dependent current signatures. These signals serve as the input to the JPA for amplification. This approach links the fundamental quantum state dynamics of the system to measurable quantities, enabling high-fidelity, real-time readout of quantum information. To calculate the current probabilities relevant, the state vectors of the system should be defined. The state vectors of the selected sub-circuit defined in Eq. 3, can be expressed as:

$$|\psi_{Q2-Q4}(t)\rangle = c_0(t)|00,0\rangle + c_1(t)|10,0\rangle + c_2(t)|01,0\rangle + c_3(t)|11,0\rangle$$
$$+ c_4(t)|00,1\rangle + c_5(t)|10,1\rangle + c_6(t)|01,1\rangle + c_7(t)|11,1\rangle \quad (4)$$

Here, $c_i$ (i = 1-7) represent the time-dependent probability amplitudes of the quantum system's basis states. In our readout scheme, probe current, $I_2$ and $I_4$, are monitored which is proportional to the probability of detecting a Cooper pair on each qubit, irrespective of the state of the other. Specifically, the currents are given by $I_2 \propto |c_5|^2 + |c_7|^2$ and $I_4 \propto |c_6|^2 + |c_7|^2$ [33-35]. These expressions reflect the population of the excited states involving qubit 2 and qubit 4, respectively. Assuming an initial state of $|00,0\rangle$ at t=0, one can solve the time-dependent Schrödinger equation to track the evolution of these probability amplitudes over time. This allows to model the resulting probe currents, which serve as the observable signals for the JPA-based readout. The time-dependent probability amplitudes are calculated as:

$$i\begin{bmatrix} \dot{c}_0(t) \\ \dot{c}_1(t) \\ \dot{c}_2(t) \\ \dot{c}_3(t) \\ \dot{c}_4(t) \\ \dot{c}_5(t) \\ \dot{c}_6(t) \\ \dot{c}_7(t) \end{bmatrix} = \begin{bmatrix} 0 & 0 & 0 & g_3 & 0 & g_1 & g_2 & g \\ 0 & \omega_1 n_1 & g_3 & 0 & g_1 & 0 & g & g_2 \\ 0 & g_3 & \omega_2 n_2 & 0 & g_2 & g & 0 & g_1 \\ g_3 & 0 & 0 & (\omega_1 n_1 + \omega_2 n_2) & g & g_2 & g_1 & 0 \\ 0 & g_1 & g_2 & g & \omega_b n_b & 0 & 0 & g \\ g_1 & 0 & g & g_2 & 0 & (\omega_b n_b + \omega_1 n_1) & g_3 & 0 \\ g_2 & g & g_1 & 0 & g_3 & 0 & (\omega_b n_b + \omega_2 n_2) & 0 \\ g & g_2 & g_1 & g_3 & 0 & 0 & 0 & (\omega_b n_b + \omega_1 n_1 + \omega_2 n_2) \end{bmatrix} \begin{bmatrix} c_0(t) \\ c_1(t) \\ c_2(t) \\ c_3(t) \\ c_4(t) \\ c_5(t) \\ c_6(t) \\ c_7(t) \end{bmatrix} \quad (5)$$

where $n_1 \equiv \langle a_1^+ a_1 \rangle$, $n_2 \equiv \langle a_2^+ a_2 \rangle$, $n_b \equiv \langle b^+ b \rangle$. The general solution as $c_i(t) = \exp[-iM*t]*c_i(0)$, where $c_i(0)$, i, and M are the state's initial condition, imagnary number, and coefficient matrix expressed in Eq. 5, can be employed to calculate $c_i(t)$. With the currents associated with $Q_2$ and $Q_4$, the time evolution of the probability amplitudes can also be used to construct the system's density matrix, providing a complete description of the quantum dynamics [33, 34]. It is important to note that the Hamiltonian given in Eq. (3) specifically describes the interaction between $Q_2$ and $Q_4$ via a shared bus resonator. However, as shown in the quantum chip configuration in Fig. 3, the system also includes two additional qubits, $Q_1$ and $Q_3$, which form a second coupled pair. Therefore, a similar approach must be applied to compute the time-dependent excitation probabilities for $Q_1$ and $Q_3$, and subsequently determine the associated currents $I_1$ and $I_3$. To enable joint readout and support coherent superposition sharing between both qubit pairs, a dedicated readout resonator is capacitively coupled to both bus resonators. The complete Hamiltonian (interaction Hamiltonian) reflecting these interactions and extended architecture is given by:

$$H_{int} = \hbar\omega_r a_r^+ a_r + \hbar\omega_b b^+ b + \hbar\omega_c c^+ c + hg_c\left(a_r^+ + a_r\right)\left(c^+ + c\right)\left(b^+ + b\right)$$

(6)

where the first term represents the energy of the readout resonator, the second and third terms correspond to the energies of the bus resonators connecting the $Q_2$–$Q_4$ and $Q_1$–$Q_3$ qubit pairs, respectively. Finally, the last term describes the capacitive coupling between these bus resonators and the shared readout resonator. Notably, Eq. 6 introduces a critical parameter, $g_c$, which plays a significant role in shaping the dynamics of the quantum circuit. It is important to emphasize that the operational frequencies of the quantum chip components—including qubits and resonators—are selected based on the JPA's gain spectrum. As such, the detuning between these elements, along with the weak-coupling regime, provides a basis for estimating the coupling strengths between qubits and their respective bus resonators, as well as the inter-bus coupling $g_c$. Furthermore, Eq. 5 presents a coupled qubits dynamics matrix whose structure is highly sensitive to these coupling strengths. Ultimately, the JPA gain profile exerts a direct influence over the chip's dynamic behavior, especially the time evolution of excitation probabilities, which determine the resulting current signals that should be amplified by the JPA. In the following section, the configuration and design methodology of the JPA is studied, explaining how its distinctive gain profile—central to the quantum chip design—is achieved. We also discuss how this gain profile inherently suppresses qubit energy leakage, thereby eliminating the need for a separate Purcell filter. Additional technical characteristics of the JPA are also explored in detail.

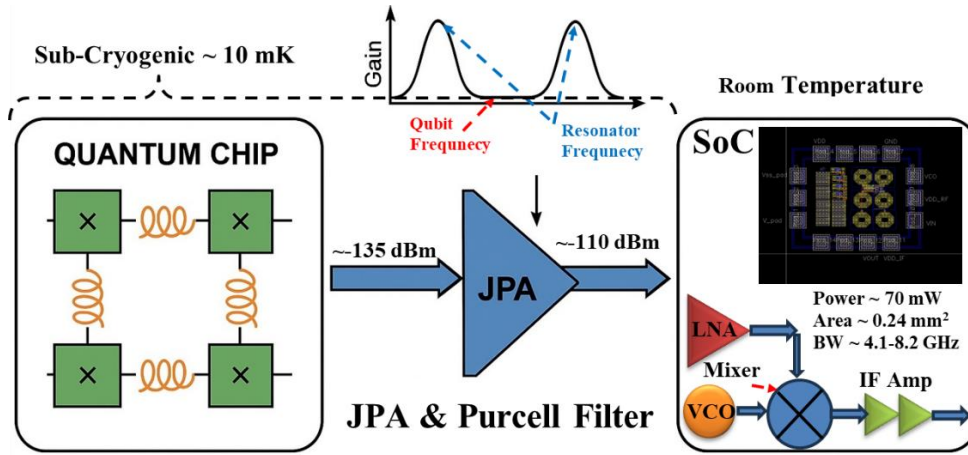

Scheme 1. Overview of the quantum readout system showing signal amplification from a superconducting quantum chip at millikelvin temperatures through a JPA with Purcell filter role involved, followed by room-temperature down-conversion and amplification using a custom SoC module.

### JPA design and Engineering

The provided schematic in Fig. 1 illustrates a sophisticated and modified version of BJPA [22], which was optimized for high linearity, wide bandwidth, and strong tunability. The structure is centered around two primary Quarton units: Quarton A and Quarton B, connected through a central resonator. Quarton A and Quarton B incorporates a primary JJ, responsible for establishing the primary nonlinear interaction and setting the core operating parameters of the amplifier. In addition, they include an array of the secondary JJs, which enhance the system's flexibility by fine-tuning the nonlinearity and gain profile. The resonator acts as a coupling bridge, enabling coherent energy exchange between the primary and secondary junctions, essential for achieving the desired gain shaping across a broad frequency range. Finally, the M cells are interconnected to form the new JPA, which is then connected to a λ/4 resonator and then capacitive coupling to other quantum circuits. Additionally, $\kappa_0$ is created due to the JPA mismatching with the circuit that capacitively coupled to it.

More technically, each Quarton includes a primary JJ or SQUID (with Josephson energy $E_{Jm}$), which plays a dominant role in defining the nonlinearity and the effective potential landscape of the cell. The layout features an array of capacitors that contribute to the overall shunt capacitance, tuning the resonant frequency and impedance matching. Also, the Quarton hosts the secondary JJ or SQUIDs (with the same Josephson energy of $E_{Js}$), indicating a weaker junction that assists in controlling the parametric gain and dynamic behavior of the amplifier. The central section, labeled as "Res.," acts as a coupling resonator between two Quartons and facilitates energy exchange between them. The schematic shows how a linear cascade of cells is interconnected to form a multi-stage amplifier to enhance gain and bandwidth. This modular approach allows for flexible design and scalability. One of the important elements, parasitic element, that can affect the performance of the system is $C_g$, which is hardly a controllable parameter in the quantum system. It affects the equivalent impedance of the quantum system that should be matched with $Z_0$, which is the $\lambda/4$ resonator intrinsic impedance. It is also assuming that the phase drop ($\varphi$) from the flux node distributed homogenously over the array (displayed in the figure); hence each SQUID senses the same phase drop [18-19, 21]. The flux variable $\phi$ represents the magnetic flux threading the SQUID loop. Its relationship with an applied flux $\phi_{ext}$ can be interpreted as phase shift in energy, in which an applied flux $\phi_{ext}$ typically manifests in terms like $cos(2\pi(\phi-\phi_{ext})/\Phi_0)$, where $\Phi_0$ is the flux quanta. This modifies the energy landscape and introduces tunability to the system.

In addition to the quantum theoretical framework used to analyze and study the proposed structure in the latter section, several simulations were conducted to validate the model and compare it with experimental outcomes [18-21]. The simulation results are presented in Fig. 2. Fig. 2a introduces the scattering parameters ($S_{11}$ and $S_{21}$) of the proposed JPA over the frequency range of 3.5–9.0 GHz. The blue curve shows the reflection coefficient ($S_{11}$), while the red curve shows the transmission ($S_{21}$). Deep dips in $S_{21}$ indicate strong resonances where energy is efficiently absorbed or transferred. The inset provides a zoomed-in view for clarity around the pumping frequency. These results demonstrate the highly resonant nature of the JPA, confirming the presence of multiple separated gain peaks essential for selective signal amplification. This is contributed to the energy transferring back and forth between Quaron A and Quarton B through the interconnection resonator. The most important result of the study is illustrated in Fig. 2b, by which the quantum chip is designed. The resonance characteristics observed in the gain spectrum fundamentally stem from the collective dynamics of the periodic nonlinear cells, back and forth energy transferring between two successive Quartons through the resonator between them. What makes this figure particularly significant is its depiction of a nonuniform, yet highly structured, gain profile extending across the C-band (approximately 4–8 GHz). The gain spectrum is marked by sharp and distinct peaks reaching up to ~24–25 dB, interspersed with deep minima approaching 0 dB. This comb-like profile indicates frequency-selective behavior, where the amplifier provides strong gain only at certain resonant modes. While this could be seen as a limitation in applications requiring broadband gain, it presents a valuable opportunity in frequency-multiplexed quantum systems, where each qubit operates at a distinct frequency. By aligning the readout frequencies with the gain peaks, one can maximize amplification efficiency while minimizing qubit's frequency at non-target frequencies to kill any qubit's energy leakage to readout circuit. We exploited this gain landscape to optimize the quantum chip design: qubit frequencies ($f_{Q1}$, $f_{Q2}$, $f_{Q3}$, $f_{Q4}$) are aligned with the valleys (gain minima) to minimize qubit energy leakage into the readout circuit; this is the main role of Purcell filter that automatically done using proposed JPA. Resonator frequencies ($f_{R1}$, $f_{R2}$, $f_{R3}$, $f_{R4}$), in contrast, are aligned with gain peaks to maximize signal amplification. Therefore, this technique mimics the functionality of a traditional Purcell filter without requiring extra components, simplifying the architecture and improving system performance. The JPA simultaneously enhances resonator signals and suppresses qubit decay through the readout channel. This novel frequency arrangement thus significantly boosts readout fidelity while preserving quantum coherence. Thus, as a main result, this engineered gain profile directly enabled a new, more integrated and efficient approach to quantum chip readout, supporting both high-fidelity measurements and complex free quantum circuits.

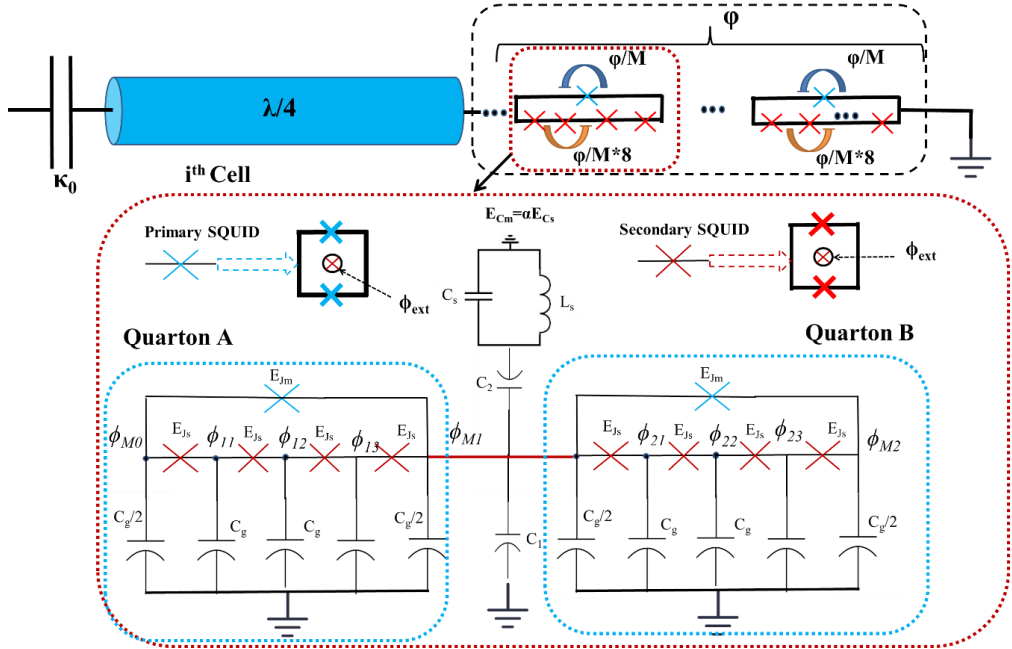

Fig. 1 Schematic of the proposed modified BJPA architecture, consisting of two coupled Quartons (A and B), each composed of an array of secondary JJ with shared parasitic capacitance $C_g$. Each Quarton is controlled via a primary SQUID through flux-biased via $\phi_{ext}$. The two Quartons are interconnected through a superconducting resonator ($\lambda/4$) to mediate energy transfer and enforce nonreciprocal dynamics. Each Quarton includes a primary JJ with energy $E_{Jm}$ and four secondary JJ with energy $E_{Js}$, couples to each other through the external resonator. The circuit's dynamic response is governed by phase bias $\varphi$, loop mutuality M, and the collective mode participation. This configuration enables both tunable gain and linearity making it suitable for scalable quantum-limited amplification over the 4–8 GHz C-band.

To support further analysis of the proposed JPA, Fig. 2c presents a comparison between the output levels of the fundamental (0,1) harmonic and the third-order (3,0) harmonic as a function of input signal power ($P_{sig}$). The red curve represents the fundamental output while the blue curve shows the third harmonic output. Across a wide range of input powers (-150 dBm to -120 dBm), the third harmonic output remains significantly suppressed relative to the fundamental. This demonstrates the JPA's good linearity and minimal harmonic distortion, critical for preserving the integrity of quantum signals. Maintaining low third-harmonic generation helps prevent unwanted mixing and signal degradation in sensitive quantum measurements. Another critical parameter in JPA design is the $P_{1dB}$, which provides a precise measure of the amplifier's linearity. Fig. 2d characterizes the gain compression behavior of the JPA by plotting gain versus input signal power at different pump bias currents. As the input signal power increases, a gradual decrease in gain occurs, indicating the device approaching saturation. Higher pump currents (e.g., 3.96 μA) maintain higher gain levels over a narrower power range compared to lower currents. The nearly flat regions show a wide dynamic range before compression. The 1-dB compression point defines the maximum input signal strength a JPA can handle without significant distortion, confirming robust amplifier performance suitable for weak quantum signals.

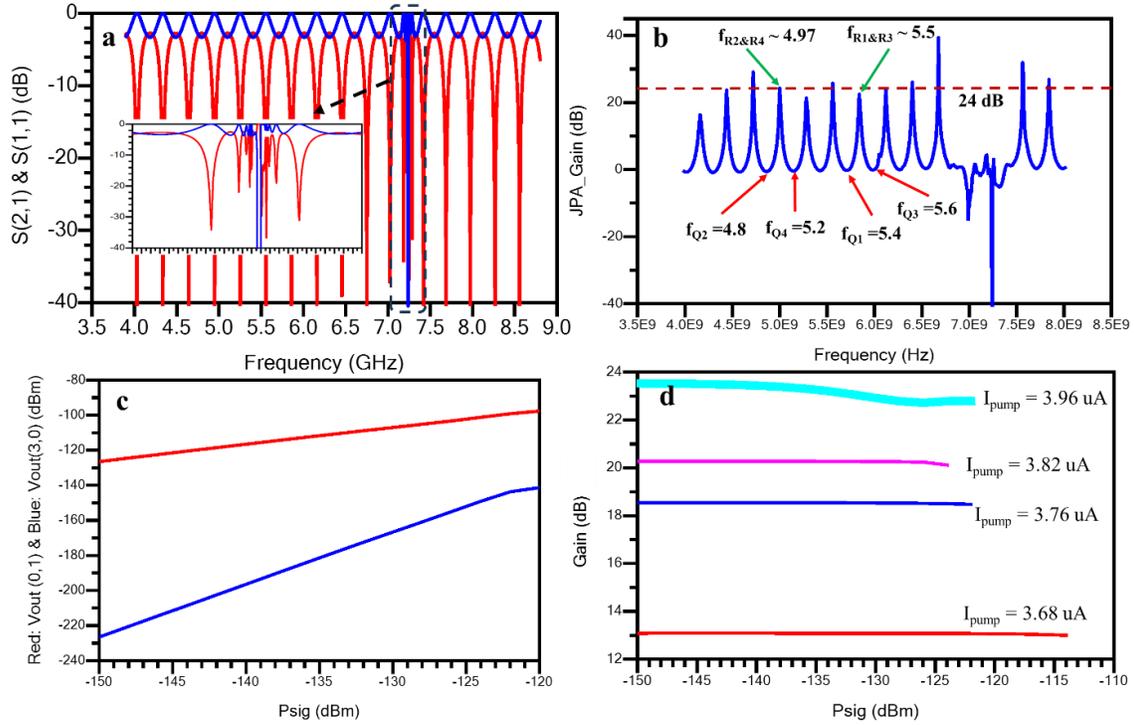

Fig. 2a) Scattering parameters (Red: $S_{11}$ and Blue: $S_{21}$) of the designed Josephson Parametric Amplifier (JPA) over the frequency range of 3.5–9.0 GHz. The inset provides a magnified view of the fine spectral features. b) Gain profile of the JPA exhibiting multiple sharp peaks (~24 dB) and deep minima (~0 dB) across the 4–8 GHz C-band. c) Comparison between the output levels of the fundamental (0,1) and third harmonic (3,0) signals versus input power. The strong suppression of third harmonics across the input power range demonstrates the high linearity of the JPA and its suitability for quantum applications. d) Gain versus input signal power at various pump currents, illustrating the 1-dB compression point behavior. Data for simulation: 8×192 JJ, $P_{sig}$ = -150 dBm, $F_{pump}$ = 7.12 GHz, $I_{pump}$ = 3.68 uA.

## Quantum chip design (coupled transmon qubits) and Engineering Approach

The four-qubit quantum chip architecture is illustrated in Fig. 3, which is designed and developed using the JPA gain profile strategy. Each qubit ($Q_1$–$Q_4$) is coupled to its own dedicated resonator ($Res_1$–$Res_4$), which connects to a central feedline allowing readout. The resonators' frequencies are aligned with the JPA's gain peaks (~24 dB) to maximize the amplification of qubit-state information, while the qubits' transition frequencies are deliberately placed at the gain minima (~0 dB) to minimize their energy leakage into the readout line. This design, derived directly from the detailed JPA characterization, emulates the protective behavior of a Purcell filter without needing extra circuitry. The quantum chip architecture along with JPA achieves Purcell-filter-like operation intrinsically through the JPA depending on the special configuration of JPA, optimizing qubit coherence and readout efficiency without additional filtering components. The structure shown in Fig. 3 and its dynamic behavior were previously analyzed using quantum theory. However, to introduce additional degrees of freedom and perform a more comprehensive analysis of the quantum chip, Qiskit Metal—a powerful and flexible design tool—is employed to thoroughly investigate the circuit. Simulations via QisKit Metal determines the resonance and cross-Kerr interactions among components by which the quantum circuit dynamics can be determined and testified. The data attained using QisKit Metal is summarized in Table. 1.

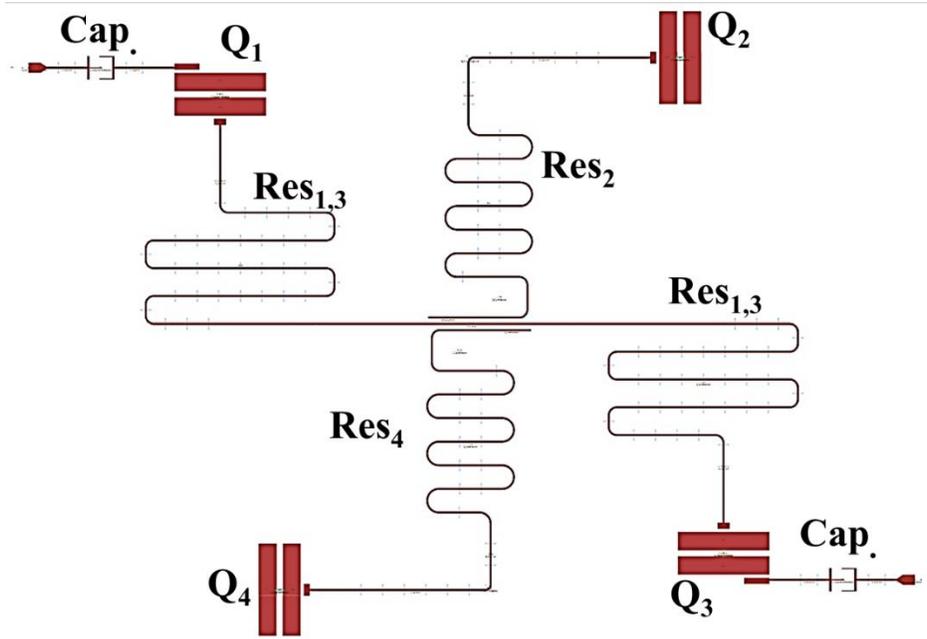

Fig. 3 Schematic layout of the designed four-qubit quantum chip based on the engineered JPA gain profile. Qubits $Q_1$–$Q_4$ are individually coupled to resonators $Res_1$–$Res_4$, which connect to a common feedline for input/output readout. Resonator frequencies are positioned near JPA gain peaks (~24 dB) to enhance signal amplification, while qubit frequencies are placed near gain minima (~0 dB) to suppress qubit energy leakage into the readout line. In this circuit, the input and output capacitors are considered just to check the effect of $Q_1$ and $Q_3$ coupling on the scattering parameters.

Table. 1 Quantum chip design using QisKit Metal and the relevant results derived from (Energy Participation Ratio) EPR analysis; Light-Blue: components Frequency; Green: qubits anharmonicity; Yellow: Resonator anharmonicity; Orange: $Q_1$-$Q_3$-$Res_{1,3}$ cross-Kerr; Dark-Blue: $Q_2$-$Q_4$-$Res_{2,4}$ cross-Kerr.

| Element | Frequency (GHz) | $Q_1$ (MHz) | $Q_2$ (MHz) | $Q_3$ (MHz) | $Q_4$ (MHz) | $Res_{1,3}$ (MHz) | $Res_2$ (MHz) | $Res_4$ (MHz) |
|---|---|---|---|---|---|---|---|---|
| $Q_1$ | 5.46 | 287 | -- | 2.34 | -- | 3.75 | -- | -- |
| $Q_2$ | 4.94 | -- | 280 | -- | 0.0018 | -- | 8.01 | 0.48 |
| $Q_3$ | 5.6 | 2.34 | -- | 289 | -- | 3.1 | -- | -- |
| $Q_4$ | 5.11 | -- | 0.0018 | -- | 257 | | 3.3 | 0.16 |
| $Res_{1,3}$ | 5.56 | 3.75 | -- | 3.1 | | 175 | -- | -- |
| $Res_2$ | 4.91 | -- | 8.01 | -- | 3.3 | -- | 154 | 0.18 |
| $Res_4$ | 4.97 | -- | 0.48 | -- | 0.16 | -- | 0.18 | 190 |

The Qiskit Metal simulation results (Table. 1) provide a detailed insight into the spectral and interaction characteristics of the 4-qubit superconducting quantum chip. Each qubit is associated with a resonance frequency carefully chosen between 4.94 GHz and 5.8 GHz to align with low-gain regions of the JPA gain spectrum, minimizing their exposure to amplified noise and spontaneous emission. Anharmonicity, defined as the deviation of energy level spacing from that of a harmonic oscillator, is essential for enabling selective qubit operations and minimizing state leakage during gate executions [10-14]. For the qubits (highlighted in green), the anharmonicity values are substantial and well-defined: $Q_1$ exhibits an anharmonicity of 287 MHz, $Q_2$ at 280 MHz, $Q_3$ at 289 MHz, and $Q_4$ at 257 MHz. These values are within an acceptable range for transmon qubits, ensuring that the $|0\rangle$ to $|1\rangle$ transition is sufficiently detuned from $|1\rangle$ to $|2\rangle$ transition. This detuning is crucial for maintaining high-fidelity gate operations and reducing leakage to higher excited states during quantum logic gates. The consistency of anharmonicity across the qubits also suggests a uniform fabrication process and effective design symmetry, both of which are favorable for scalable quantum architectures.

Cross-Kerr coupling values, expressed in MHz, reflect the nonlinear dispersive interaction between components, revealing how shifts in one element's state influence the others. For example, $Q_1$ and $Q_3$ show mutual interactions through $Res_1$ and $Res_3$, as indicated by a non-negligible cross-Kerr (~2.3 MHz to 8 MHz), supporting coupling operations. Accordingly, $Q_2$ and $Q_4$ interact similarly through $Res_2$ and $Res_4$ with coupling strengths ranging from 0.16 to 3.3 MHz. In contrast, interactions between non-coupled qubit pairs (e.g., $Q_1$–$Q_4$, $Q_2$–$Q_3$) are minimal, but not zero; it is important to note that Qiskit Metal's EPR analysis is limited to handling a maximum of four modes. Consequently, the interaction between $Q_1$ and $Q_4$, for example, is not captured in the table. The resonator frequencies—ranging from 4.91 GHz to 5.56 GHz—are matched to the high-gain peaks of the JPA (Fig. 2b), achieving up to 24 dB amplification. These high-gain resonator channels ensure that readout signals from quantum chip are strongly amplified, while qubits energy leakage is suppressed severely, improving signal-to-noise ratio during measurement. The Qiskit Metal-derived data verifies that the quantum chip's spectral design supports both robust coupling for computation and minimal exposure to loss channels. This reflects a high degree of quantum control, precise fabrication tuning, and simulation-informed engineering for scalable quantum processing.

Finally, using the data extracted from Qiskit Metal, we proceed to analyze the dynamics of the quantum chip. From latter section, Eq. 3 through Eq. 6 provide a comprehensive quantum mechanical framework for studying the circuit. For simplicity, we consider only half of the chip, coupling between $Q_2$ and $Q_4$ via resonators, under the assumption of circuit symmetry. By solving the Lindblad master equation and employing Eq. 5, the current-related probabilities $I_2$ and $I_4$ (and similarly $I_1$ and $I_3$) is computed. This modeling approach can be regarded as a standard framework, as the essential parameters—qubit and resonator frequencies, as well as coupling strengths such as cross-Kerr coefficients—are directly obtained from Qiskit Metal simulations. Additionally, we aim to validate the quantum theoretical model by comparing its predictions with experimental measurements. In this study, we examine how varying the coupling strength $g$ between $Q_1$, $Q_3$, and their respective resonators influences the current probabilities. This parameter is intentionally selected due to its role in establishing quantum correlations such as entanglement within the chip. As shown in Fig. 4a, under weak coupling conditions, the current probabilities exhibit clean sinusoidal oscillations. However, increasing the coupling strength $g$, as seen in Fig. 4b, leads to deviations from sinusoidal behavior, resulting in more complex, mixed waveform dynamics. This transformation aligns with experimental findings in [33, 34], where similar current patterns were observed as entanglement emerged in experimental circuits. To further clarify this behavior, the frequency spectra of the results in Fig. 4c and Fig. 4d is compared. While both cases show two primary frequencies, $f_0$ and $f_1$, the stronger coupling case (Fig. 4d) reveals the emergence of an additional frequency component, $f_c$, which corresponds to the onset of entangled states in the system. As an interesting conclusion, to sense the entangled signals generated through the quantum chip, the readout resonator frequency should be tuned with $f_c$. In line with, Qiskit Metal simulations confirm the presence of a mode—denoted as $f_c$—which facilitates coupling between $Q_2$, $Q_4$ (or $Q_1$, $Q_3$), and the intermediate resonators. This directly supports the last term in Eq. 3 or Eq. 6, where the coupling strength $g$ governs the creation and control of entanglement within the quantum chip. In the following, we attempt to illustrate how the modes in the quantum chip are distributed when g is arisen. As illustrated in Fig. 5, the mode $f_c$ acts as a shared channel that links $Q_1$, $Q_3$, and the resonator according to the Hamiltonian presented in Eq. 3, enabling tunable interaction. Thus, by adjusting the coupling parameter, $g$, one can effectively manipulate the entanglement between qubits via the mediated resonator mode in the same way with the results illustrated in Fig. 4.

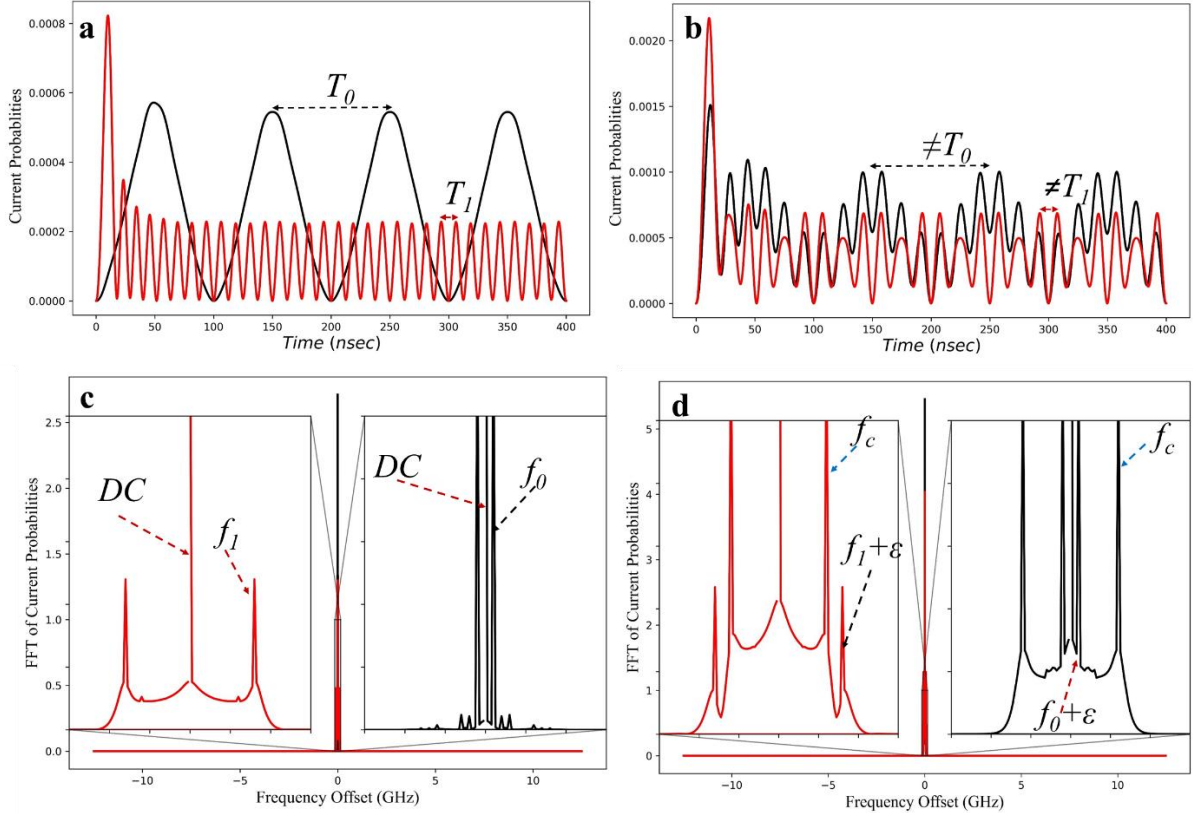

Fig. 4 Time evolution (a,b) and FFT (c,d) of the Current Probabilities ($I_1$: Black, $I_2$: Red) for different coupling factor among $Q_1$, $Q_3$, and $Res_{1,3}$; $f_{Q1}, f_{Q3}$, and $f_{res1,3}$ are selected from Table .1, and drive frequency $\omega_d = 5.5$ GHz, $g_1 = 0.8$ MHz, $g_2 = 3.7$ MHz, $g_3 = 2.3$ MHz; a) and c) $g = 0.2*g_3$, b) and d) $g = 1.8*g_3$; DC in Fig. c indicates the Fourier series DC term; inset figures in c) and d) are used to clear up the frequency offset close to zero.

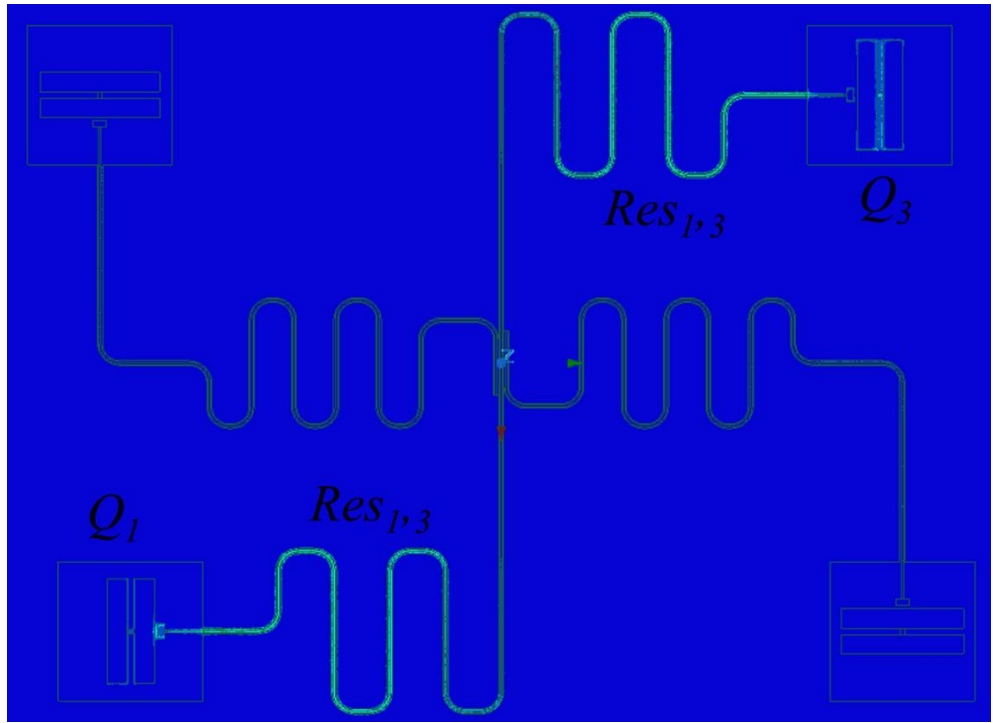

Fig. 5 Entangled mode in the designed quantum chip using the Qiskit Metal simulation; coupling among $Q_1$, $Q_3$, and the resonator, by which the existence of $g$ factor in the Hamiltonian is confirmed.

Up to this point, we have demonstrated that an effective quantum chip can be designed by leveraging the unique gain profile of a newly proposed JPA. This tailored gain spectrum eliminates the need for an additional Purcell filter, simplifying the overall architecture. Furthermore, it has been shown that the JPA is capable of amplifying quantum signals to a level suitable for detection by analog circuits such as a HEMT amplifier or directly by a SoC receiver. Consequently, the next section shifts focus to the design of a SoC circuit aimed at sensing the output signals from the quantum chip, amplifying and filtering them, and generating clean sinusoidal waveforms ready for digitization and further processing.

**Quantum Analog chip: SoC design using 45 nm CMOS**
The presented SoC block diagram (Fig. 6) highlights an integrated analog receiver architecture designed for ultra-low noise and frequency down-conversion applications. The schematic shows the major components of the receiver, including an LNA, Bandpass Filter (BPF), RF Amplifier, VCO, Mixer, IF amplifiers, and Lowpass Filters (LPFs). Each element's power consumption (PC) is annotated to emphasize the importance of power budgeting—a critical consideration in SoC in which heat dissipation is severely limited. The LNA, operating over a 4.1–8.2 GHz bandwidth, delivers high gain (~72 dB) with low noise (NF < 1 dB) and consumes ~28.2 mW. The VCO is more energy-efficient, consuming ~0.18 mW. The VCO operates across the same RF bandwidth and generates local oscillator signals (e.g., 5.0 GHz) for frequency conversion in the mixer. The mixer itself draws ~1.6 mW. The IF chain, responsible for filtering and amplifying down-converted signals below 250 MHz, includes an amplification stage with a gain of ~32 dB and power consumption of ~9.5 mW for each IF. Altogether, the design maintains a power-efficient profile (~70 mW total), which is crucial for scalable integration in quantum computing systems. Notably, each block refers to a layout-based implementation (post-layout) rather than an ideal schematic, ensuring practical modeling for real SoC realization.

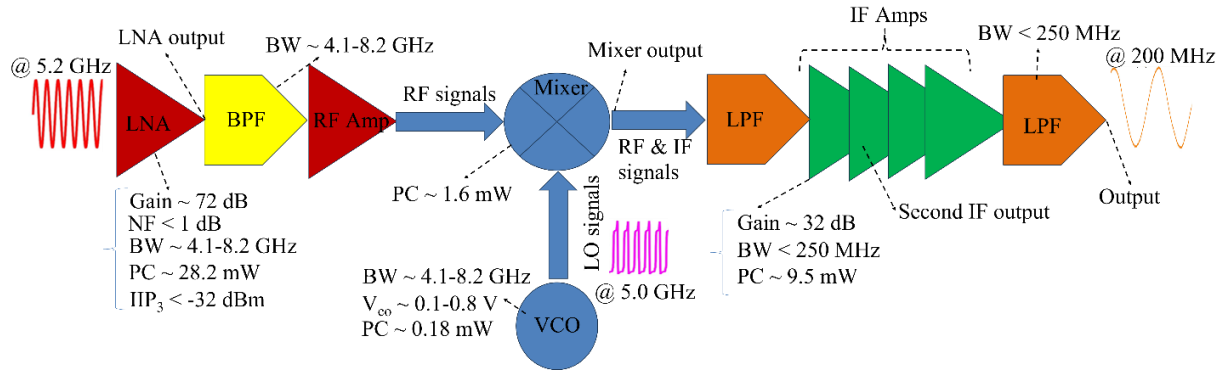

Fig. 6 Layout-based SoC analog receiver chain and post-layout transient simulation results. The SoC symbol diagram showing the cascade of LNA, VCO, Mixer, and IF amplifiers; this schematic attempt to conveys some specific characterizations and a few relevant power consumptions (PC) for each critical element used to determine the SoC power budget.

In line with the SoC component's simulation results, Fig. 7 presents the gain and NF performance of the LNA designed for the proposed SoC-based quantum receiver, targeting operation within the C-band (4–8 GHz). In Fig. 7a, the NF curve remains well below the 1 dB threshold across the entire band, reaching a minimum of approximately 0.94 dB near 5.6 GHz. This indicates that the LNA introduces minimal additional noise to the input signal, which is critical for detecting extremely weak quantum signals, typically in the nanovolt range. Fig. 7b shows that the amplifier achieves a nearly flat gain profile around 72 dB throughout the C-band, with only minor variations. The flatness and high gain ensure that the signal is sufficiently amplified for further processing stages without distortion or bandwidth-dependent gain fluctuations. This level of performance demonstrates the suitability of the LNA for integration into low-noise, high-fidelity quantum readout architectures. Thus, the LNA meets the design objectives of sub-1 dB noise figure and high, flat gain in the 4–8 GHz band, supporting reliable signal acquisition and down-conversion in quantum systems. The combination of very high gain and a low NF suggests that the use of HEMT amplifiers in the signal chain may be unnecessary. This is particularly significant considering that, unlike HEMT technology, CMOS can be integrated on the same substrate as the quantum chip, offering advantages in scalability and system integration.

The post-layout transient simulation results of the SoC receiver are shown in Fig. 8. The input waveform represents the input RF signal applied to the SoC, centered around 5.2 GHz. The voltage level is extremely low (~100 nV is a typical value), indicating the necessity of a very low-noise environment and sensitive front-end amplification. At this stage, noise effects dominate over signal amplitude, necessitating robust LNA performance immediately after this point. After passing through the LNA, the input signal shows significant amplification, reaching amplitudes in the range of several hundred microvolts (μV). Importantly, the frequency remains locked at 5.2 GHz, ensuring that the LNA introduces minimal distortion. The LNA amplifies the desired signal while preserving its spectral purity, preparing the signal for effective mixing in the subsequent stage. This amplified signal is crucial for achieving a good signal-to-noise ratio (SNR) before frequency conversion. The VCO generates a strong, stable sinusoidal signal also at 5.2 GHz with a sensible amplitude (around several millivolts). Notably, the VCO signal must be very pure (low phase noise), and its steady, clean waveform in this result confirms proper design and layout isolation strategies. At the mixer output, two dominant frequency components appear: the sum (10.4 GHz) and difference (near

DC, a few hundred MHz). The result is a complex envelope containing both high-frequency RF and low-frequency IF components. The visible beat pattern and amplitude modulation in the waveform show successful mixing action. However, filtering and additional amplification are needed to isolate the desired 200 MHz IF signal from this combination of frequencies. The output of the second IF shows a superposition of the RF (residual 5.2 GHz) and the emerging IF (200 MHz) components. The envelope grows smoother compared to previous stages, demonstrating that the unwanted high-frequency contents are progressively suppressed. It indicates the effectiveness of the $IF_1$ and $IF_2$ amplifiers, which prioritize gain at the target IF frequency while attenuating residual RF signals. At the final output, the waveform becomes a clean, large-amplitude 200 MHz sinusoid. The successful suppression of the original 5.2 GHz carrier is evident, indicating that the IF chain ($IF_1$ to $IF_4$) amplifies and filters effectively. The signal amplitude is significantly higher (~20 mV P-P), enabling further processing like analog-to-digital conversion (ADC) with high fidelity. This final output validates the full functionality of the SoC receiver chain from signal reception to frequency down-conversion. The floor plane of the SoC RF front-end as a complete layout is illustrated in Fig. 9, designed for high-performance signal detection within the 4.1–8.2 GHz C-band spectrum. The chip integrates multiple critical sub-blocks, including a LNA, VCO, Mixer, IF amplifier & LPF, coupling capacitors (~3 pF), and RF matching networks. The total chip area is approximately 0.6 mm×0.4 mm, reflecting a compact and highly integrated design suitable for quantum applications where die area and signal integrity are critical. The LNA is the only component utilizing spiral inductors, strategically employed to achieve high gain while maintaining a NF below 1 dB.

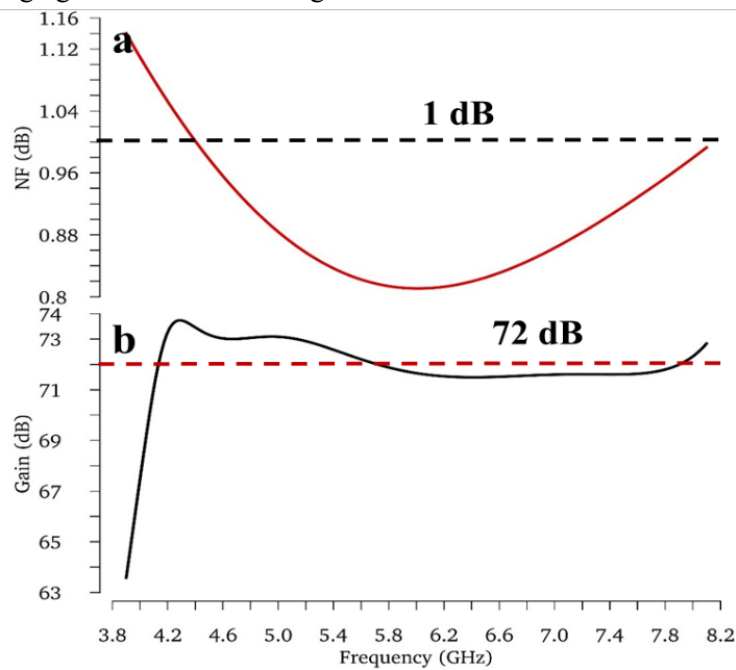

Fig. 7a) LNA NF (dB) and b) gain (dB) vs. Frequency (GHz).

To minimize chip area and parasitic effects, the rest of the SoC emphasizes active components over bulky passive ones. The use of active components, suitable impedance matching, and feedback networks ensures efficient input-output matching and helps reduce reflection loss, which is essential for maximizing gain-bandwidth product and system linearity. To further optimize the noise figure, feedback-based noise cancellation techniques are incorporated. Moreover, layout-level engineering emphasizes symmetry and isolation, which helps suppress coupling-induced parasitics and enhances overall RF performance. The

VCO in the current design is free-running and implemented directly on-chip. It generates a stable sinusoidal signal with low phase noise, sufficient for mixing without significantly degrading the noise figure. There is no external phase locked loop (PLL) [23-27] in this version of SoC. Additionally, the current layout focuses on analog RF signal detection and processing. In other words, digital control components—such as biasing logic or PLL loop filters—are not included in this version, but they can be integrated if a PLL is incorporated in future iterations. On-chip biasing is achieved through dedicated pads and biasing resistors, and tuning voltages for VCO are externally programmable. The design is finally tuned to generate a stable ~200 MHz sinusoidal output (recommended IF), suitable for direct digitization or further processing. This is chosen to balance trade-offs between filtering ease, minimal LO-RF leakage, and compatibility with room-temperature analog to digital converter (ADC). The IF chain (LPF included) includes four amplifiers optimized for <250 MHz with a combined gain of ~32 dB and very low distortion, delivering a clean sinusoidal output suitable for digitization.

One of the key performance goals in the design is low power consumption, successfully stabilized around 70 mW, making the SoC favorable for battery-powered or cryogenic environments. An important outcome of the design is that post-layout simulation results demonstrate minimal degradation due to parasitic elements, confirming that layout parasitics have been well controlled. This is achieved through meticulous routing, minimal metal crossover, and grounding strategies. The compact form factor, high integration, and robust performance across the full C-band make this SoC a strong candidate for scalable quantum receivers and high-sensitivity RF applications. However, one of the interesting and important consideration in quantum-RF circuit design is estimating the number of qubits that the proposed receiver can support, which can be approximated by the ratio: Number of qubits = (Receiver bandwidth) / (Bandwidth required per qubit). The SoC receiver demonstrates a 4 GHz bandwidth while consuming 70 mW of power. In a traditional way, assuming a conservative 10 MHz bandwidth per resonator—suitable for fast qubit readout before decoherence—and a 15 MHz spacing between adjacent qubits, the design supports readout for up to 160 qubits. This highlights the receiver's scalability and efficiency, achieving an impressive power consumption of only 0.44 mW per qubit. However, if the bandwidth is allocated according to the present design—where each pair of coupled qubits and their associated resonators require approximately 400 MHz—the SoC can individually support up to 20 qubits and their respective resonators. Additionally, the proposed architecture achieves a power consumption of around 3.5 mW per qubit.

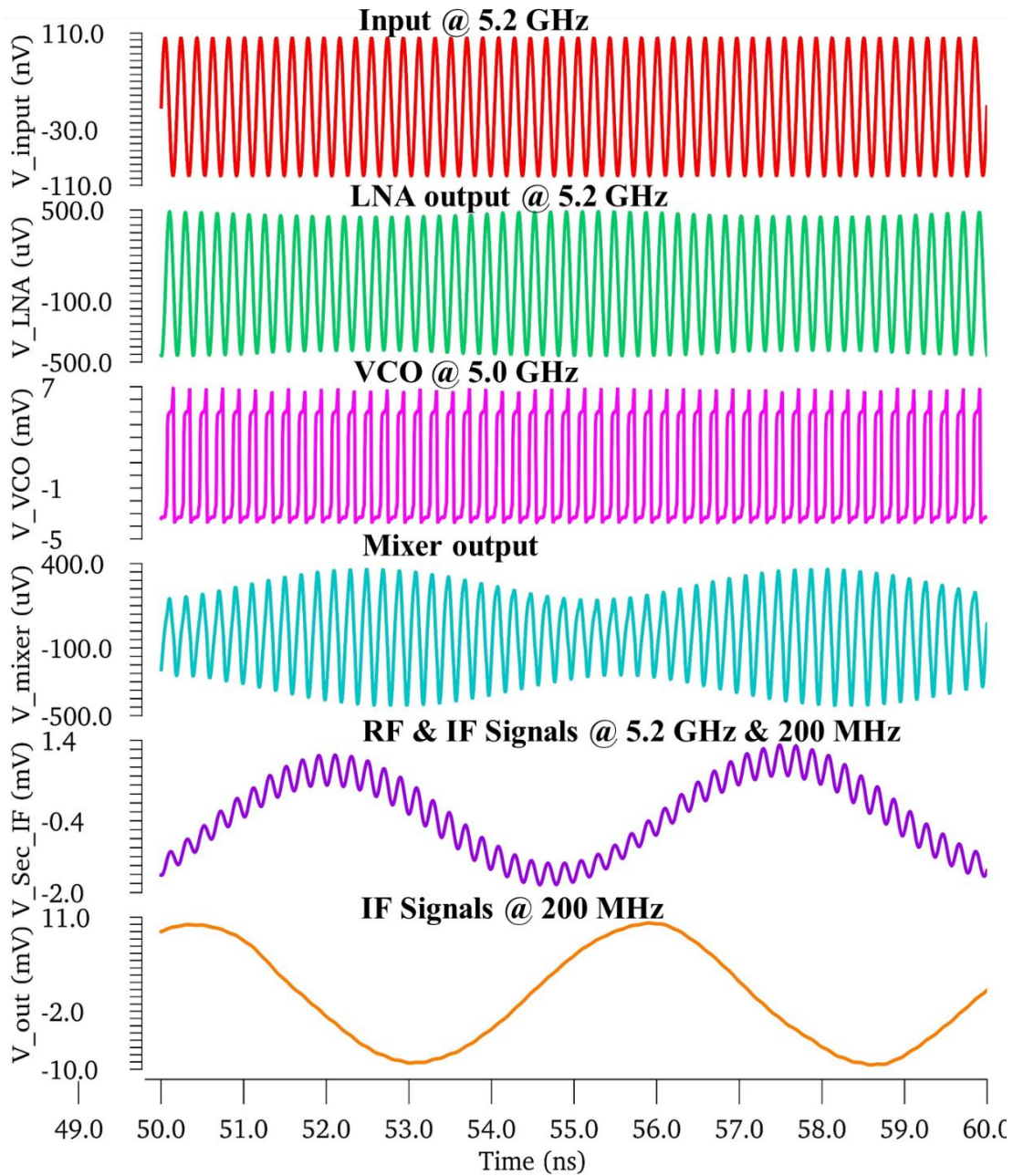

Fig. 8 Post-layout transient simulation results of the SoC receiver. The input 5.2 GHz signal with an amplitude of ~100 nV is progressively amplified by the LNA, mixed with the VCO output, and down-converted to a 200 MHz IF signal. The waveforms illustrate the signal evolution at key stages: input, LNA output, VCO output, mixer output, combined RF-IF signal, and final amplified IF output.

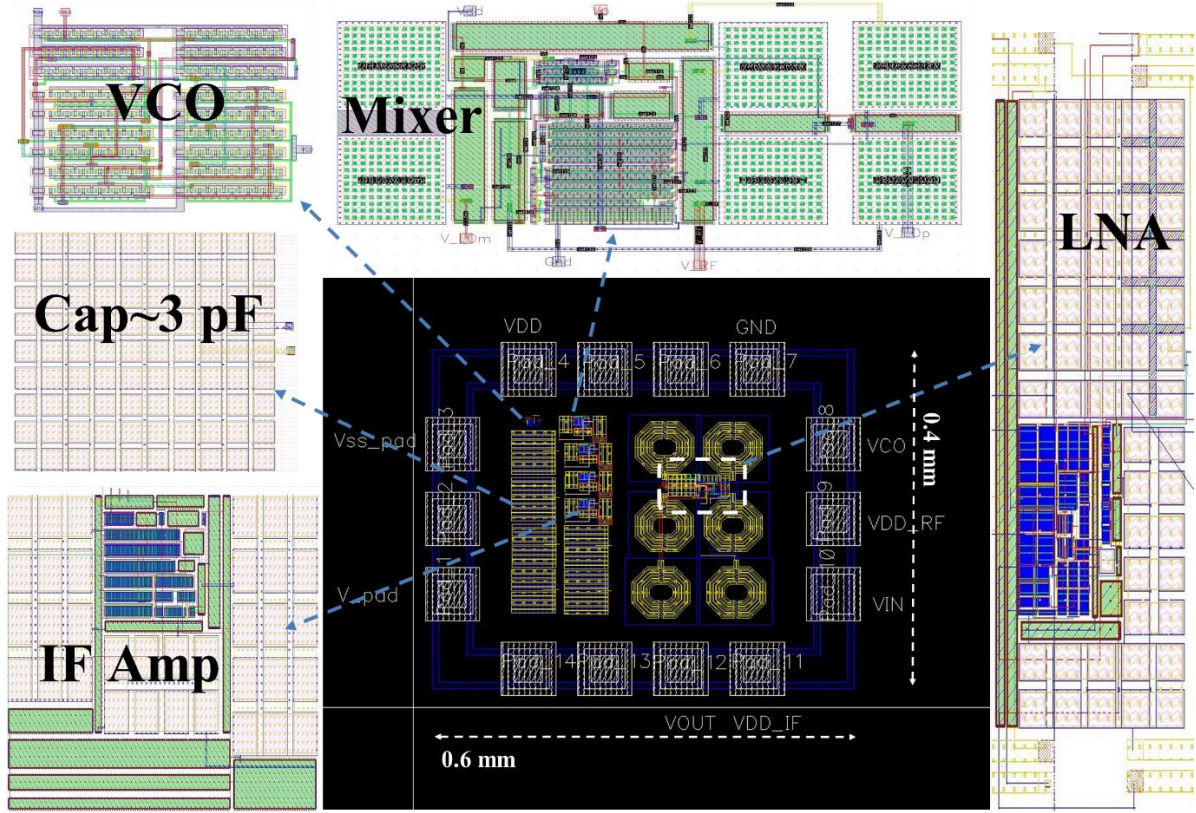

Fig. 9 Floor plane of the fully integrated SoC designed for C-band (4.1–8.2 GHz) quantum-RF applications. The SoC includes a LNA, Mixer, VCO, IF amplifier, coupling capacitors (~3 pF), and RF filters. The total chip size is approximately 0.6×0.4 mm². The design emphasizes compactness by minimizing the use of inductors (only in LNA), employing active components for space efficiency, implementing noise cancellation and feedback techniques, and using optimized matching networks to reduce the noise figure. Post-layout simulations confirm that layout parasitics have a negligible impact on performance.

**Conclusions**

In this work, in one side, we introduced a co-designed architecture that combines a novel JPA with a quantum chip to achieve efficient and scalable quantum signal readout. The JPA was designed and engineered with a tailored gain profile that not only provides quantum-limited amplification but also intrinsically suppresses qubit energy leakage—eliminating the need for a conventional Purcell filter. By aligning qubit frequencies with gain minima and resonator frequencies with gain peaks, we achieved selective amplification and reduced quantum circuit complexity. The four-qubit superconducting quantum chip was designed using Qiskit Metal and analyzed using full quantum theoretical methods to ensure accuracy in modeling coupled dynamics, cross-Kerr interactions, and signal generation. Through quantum theoretical analysis, it was demonstrated that the current probabilities arising from the coupled qubits can exhibit entanglement. To accurately detect this entangled behavior via the readout circuit, its resonance frequency must be tuned to match the entanglement frequency. In the other side, the analog SoC developed in 45 nm CMOS technology was designed and optimized for low power (~70 mW), wide bandwidth (4.1–8.2 GHz), and minimal noise (NF < 1 dB). Post-layout simulations confirm robust performance with negligible parasitic degradation. This integrated architecture supports readout for up to 20 coupled qubits with ~3.5 mW per qubit, demonstrating a scalable, compact, and cryogenically efficient solution. As an interesting point, the proposed design represents a significant step forward in quantum readout technology, offering a practical platform for next-generation quantum computing systems.